 \documentstyle[amsfonts,multicol,fancyheadings,eqsecnum,prd,aps]%
 {revtex}

\def\buildchar#1#2#3{\null \! \mathop {\vphantom {#1}\smash
#1}\limits ^{#2}_{#3}\!\null }
\def\OT#1{\buildchar{{#1}}{\;_\sim}{}\/}
\def\UT#1{\buildchar{{#1}}{}{^\sim}\/}
\def\OTT#1{\buildchar{{#1}}{\;_\approx}{}\/}

\begin{document}
\draft


\title{Gauge symmetries in Ashtekar's
    formulation of general relativity}

\author{D.\ C.\ Salisbury 
\footnote[2]{Electronic address: dsalisbury@austinc.edu}} 
\address{Department of Physics, 
Austin College, Sherman, Texas 75090-4440, USA}

\author{J.\ M.\ Pons
\footnote[1]{Electronic address: pons@ecm.ub.es}}
\address{Departament d'Estructura i Constituents de la Mat\`eria, 
Universitat de Barcelona,\\
and Institut de F\'\i sica d'Altes Energies,\\
Diagonal 647, E-08028 Barcelona, Catalonia, Spain}

\author{L.\ C.\ Shepley 
\footnote[3]{Electronic address: larry@helmholtz.ph.utexas.edu}} 
\address{Center for Relativity, Physics Department, \\
The University of Texas, Austin, Texas 78712-1081, USA \\ ~}

\maketitle

\begin{abstract}
It might seem that a choice of a time coordinate in Hamiltonian
formulations of general relativity breaks the full
four-dimensional diffeomorphism covariance of the theory. This is
not the case. We construct explicitly the complete set of gauge
generators for Ashtekar's formulation of canonical gravity. The
requirement of projectability of the Legendre map from
configuration-velocity space to phase space renders the symmetry
group a gauge transformation group on configuration-velocity
variables. Yet there is a sense in which the full four-dimensional
diffeomorphism group survives. Symmetry generators serve as
Hamiltonians on members of equivalence classes of solutions of
Einstein's equations and are thus intimately related to the
so-called ``problem of time" in an eventual quantum theory of
gravity.
\end{abstract}

\pacs{04.20.Fy, 11.10.Ef, 11.15.-q 
\hfill gr-qc/0004013}

\maketitle

\section{INTRODUCTION AND QUANTUM MOTIVATION}

The symmetry of four-dimensional spacetime diffeomorphisms lies at
the conceptual core of Einstein's theory of classical general
relativity. Any viable quantum theory of gravity must recognize
and preserve this symmetry, at least in an appropriate
semi-classical regime.  In a recent series of papers we have shown
that the full spacetime diffeomorphism group symmetry is present
in phase space (cotangent bundle) versions of conventional general
relativity \cite{pons/salisbury/shepley/97}, in
Einstein-Yang-Mills theory \cite{pons/salisbury/shepley/99a}, and
in both real triad \cite{pons/salisbury/shepley/99b} and complex
Ashtekar formulations of gravitation
\cite{pons/salisbury/shepley/99c}.  We constructed both
infinitesmal and finite canonical gauge symmetry generators in a
phase space which includes the gauge variables of the models.
Rigid time translation is, however, not a gauge symmetry in phase
space, and we shall begin to explore some of the profound
implications of this fact below in the context of the Ashtekar
loop approach to quantum gravity.

Foremost among current conceptual and technical problems with
theories of quantum gravity is the ``problem of time''. Time
evolution and spacetime diffeomorphisms are inextricably linked;
every spacetime diffeomorphism generator is, in a sense to be
explained below, a generator of time evolution. Consequently the
symmetries we display in this paper have a direct bearing on
several aspects of the problem of time. Let us first focus on the
implications of a choice of time foliation in the classical
theory. Some authors have suggested that since the canonical
gravitational Hamiltonian vanishes there is no time evolution in
quantum gravity; time is said to be ``frozen''\cite{isham/92}.
Since rigid time translation is after all a symmetry in the
Lagrangian formalism, these authors observe that we should not be
dismayed with this fact. Since a foliation translates into a gauge
choice in the quantum theory we need to inquire into the relation
between gauge choices. H\'aj\'\i{}\v{c}ek has shown in some simple
cases that the canonical quantization procedure leads to unitarily
inequivalent representations \cite{hajicek/99}. Perhaps an even
more disquieting consequence of a time foliation in the canonical
approach is the resulting either real or apparent quantum
spatio-temporal asymmetry. To date only spatial discreteness (in
area and volume) has emerged in the loop approach
\cite{rovelli/smolin/95}. And we have no prescription for
transforming to a new time slice.

Finally, we seem to have no means of predicting the outcome of
what must surely be one of the most basic thought experiments in
quantum gravity: what is the spacetime separation between two
timelike separated events?  We could imagine that these events
could be characterized, for example, by ambient matter.  Surely we
would in general expect a range of outcomes.  No such quantum
fluctuations arise in the current canonical approaches to quantum
gravity.

\section{PROJECTABILITY OF SYMMETRY VARIATIONS UNDER THE LEGENDRE MAP}

All of these difficulties stem from efforts at excising gauge
variables from the quantum theory.  But these gauge variables play
a fundamental role in the classical Hamiltonian symmetry
structure.  We have investigated conditions that must be fulfilled
by gauge symmetry transformations in the original Lagrangian
formalism which can be mapped under the Legendre map to phase
space.  (More precisely we require that phase space gauge
variation pullbacks be configuration-velocity space variations.)

The Lagrangian density for vacuum gravity, where the configuration
variables are the metric
\begin{equation}
    (g_{\mu\nu}) = \left(
        \begin{array}{cc}
            -N^{2}+N^{c}N^{d}g_{cd} & g_{ac}N^{c}  \\
            g_{bd}N^{d} & g_{ab}
        \end{array}
                \right) \ ,
    \label{eq:gmunu}
\end{equation}
does not depend on the time derivatives of the lapse $N$ and shift
functions $N^a$ ($\mu, \nu$ are spacetime indices; latin indices
from the beginning of the alphabet are spatial indices).
Therefore projectable variations may not depend on these time
derivatives\cite{pons/salisbury/shepley/97}.

The projectable infinitesmal spacetime diffeomorphisms in
conventional gravity are of the form $x'^\mu = x^\mu
-\epsilon^\mu$ where  the descriptor $\epsilon^\mu$ contains a
compulsory lapse and shift dependence and $\xi^\mu$ is an
arbitrary function:
\begin{equation}
    \epsilon^\mu = \delta^\mu_a \xi^a + n^\mu \xi^0\ .
    \label{eq:epsilon}
\end{equation}
The normal $ n^\mu$ to the fixed time hypersurface is expressed as
follows in terms of the lapse and shift:  $n^\mu =
(N^{-1},-N^{-1}N^a)$.

If gauge symmetries exist beyond those induced by diffeomorphisms
one obtains additional projectability conditions.  We have shown
that in a real triad approach to gravity in which the
configuration variables are taken to be a densitized triad $ \OT
T^{a}_{i} := t T^{a}_{i}$ and a gauge function $\Omega^i_0$,
projectable variations must also not depend on time derivatives of
$\Omega^i_0$\cite{pons/salisbury/shepley/99b}: $t$ is the
determinant of the covariant triad $t^i_a$, from which one forms
the 3-metric $g_{ab}=t^i_a t^j_b \delta_{ij}$.  Over- and
undertildes label the integer weight of the density under spatial
diffeomorphisms.  Latin indices from the middle of the alphabet
range from $1$ to $3$ and label the triad vectors.  These indices
are raised and lowered with the Kronecker delta.  $T^{a}_{i}$ is
the inverse triad to $t^{i}_{a}$.  It turns out then that
spacetime diffeomorphism-induced gauge variations are not by
themselves projectable; an $SO(3,R)$ triad rotation fixed by the
arbitrary function $\xi^0$ must be added to them.  The
infinitesmal descriptor of the required rotation is $\xi^i =
\Omega_\mu^i n^\mu \xi^0$ where $\Omega_a^{ij} = \epsilon^{ijk}
\Omega_a^k$ are the 3-dimensional Ricci rotation coefficients.
(The infinitesmal $SO(3,R)$ variation of a triad vector
corresponding to a descriptor $\xi^i$ is $\delta_R(\xi^i)t^i_a =
-\epsilon^{ijk} \xi^j t^k_a$.  $\Omega^i_\mu$ transforms as a
spacetime connection: $\delta_R(\xi^i) \Omega^i_\mu =
-\xi^i_{,\mu} -\epsilon^{ijk}\xi^j \Omega^k_\mu$.  )

These triad variables are in fact among the set of configuration
variables in Ashtekar's complex connection approach to general
relativity. The connection is formed with the 4-dimensional Ricci
rotation coefficients $\Omega^{IJ}_\mu$:
\begin{equation}
    A^i_\mu =\epsilon^{ijk} \Omega^{jk}_\mu + i \Omega^{0i}_\mu\ .
    \label{eq:Ai}
\end{equation}
(Indices $I$,$J$ range from $0$ to $3$ and are tetrad labels.)
Since the action is independent of the time derivatives of the
connection components $A^i_0$, projectable symmetry variations
must be independent of this time derivative. Thus it turns out
once again that in order to be projectable, variations induced by
infinitesmal spacetime diffeomorphisms, which already require the
same lapse and shift dependence as above, must be accompanied in
general by $SO(3,C)$ triad rotations
\cite{pons/salisbury/shepley/99c}. The functional form of the
required infinitesmal descriptor, $\xi^i = A_\mu^i n^\mu \xi^0 -i
N^{-1} T^{ai} N_{,a} \xi^0$, differs from the real triad case, but
the required rotations of course agree in the real triad sector of
the Ashtekar theory.

\section{SYMMETRY GENERATORS}

Phase space in the Ashtekar theory is coordinatized by the
canonical pairs $\{\OT T^{a}_{i}, i A^i_a\}$, plus the gauge
functions $\{\UT N, N^a , -A^i_0\}=: N^A$, with their canonical
momenta, which are primary constraints: $\{P,P_a,-P_i\}=: P_A$.
The physical phase space is further constrained by secondary
constraints $\{\OTT{\cal  H}_0,\OT{\cal  H}_a,\OT{\cal
H}_i\}=:{\cal H}_A$. These constraints generate symmetry
variations of the non-gauge variables. The complete generators
(complete in the sense that they also generate variations of the
gauge variables), with infinitesmal descriptors $\{\UT
\xi^0,\xi^a, \xi^i \} =: \xi^A$, are of the form
\begin{equation}
    G[\xi] = P_{A} \dot\xi^{A}
       + ({\cal H}_{A}
       + P_{C''}N^{B'}{\cal C}^{C''}_{AB'})\xi^{A}\ ,
           \label{GGGGG}
\end{equation}
where the structure functions are obtained from the closed Poisson
bracket algebra
\begin{equation}
    \{ {\cal H}_{A},{\cal H}_{B'} \}
    =: {\cal C}^{C''}_{AB'} {\cal H}_{C''}\ ,
\end{equation}
and where spatial integrations over corresponding repeated capital
indices are assumed.

\section{RIGID TIME TRANSLATION AND A QUANTUM PROPOSAL}

We observe that $G[\xi] \delta t$ effects rigid time translations
on those solution trajectories satisfying
\begin{eqnarray}
N &=& t \UT \xi^0, \nonumber\\ N^a &=&\xi^a, \nonumber\\ -A^i_0 +
A^i_a N^a &=& \xi^i, \label{N}
\end{eqnarray}
where the descriptors $\xi^{A}$ are here taken to be finite. Thus
every generator $G[\xi] \delta t$ with non-vanishing and positive
$\UT \xi^0$ is in this sense a generator of time evolution. Of
course, on solutions whose gauge functions are not related to the
descriptors as in (\ref{N}) the engendered variation is more
general.

The general finite generator is
\begin{equation}
    {\cal T} \exp\left(\int^{t_{0}+\tau}_{t_{0}} dt \,
                    \{ - , G[\xi] \}\right)\ ,
    \label{eq:calT}
\end{equation}
where ${\cal T}$ is the time ordering operator.  This suggests a
tentative implementation of this much larger symmetry in quantum
gravity. We propose to retain the gauge variables in an expanded
loop structure. In so doing we will be able to construct true
spacetime holonomies (with the full spacetime connection
$A^i_\mu$), and since the lapse and shift will constitute quantum
operators, quantum fluctuations in the full spacetime metric will
emerge. Physical states can be constructed in principle in this
formalism by integrating out the full spacetime diffeomorphism
gauge freedom, generalizing an expression proposed by Rovelli
\cite{rovelli/98}; we propose a functional integral projector onto
physical states of the form
\begin{equation}
    {\cal T}\left( [D\xi]e^{-i\int dt G[\xi]}\right)\ .
    \label{eq:path}
\end{equation}


\end{document}